# Polylactic acid-nanocrystalline carbonated hydroxyapatite (PLA-cHAP) composite: preparation and surface topographical structuring with direct laser writing (DLW)

Edita Garskaite[1], Laurynas Alinauskas[1], Marian Drienovsky[2], Jozef Krajcovic[2], Roman Cicka[2], Marian Palcut[2], Linas Jonusauskas[3], Mangirdas Malinauskas[3], Zivile Stankeviciute[1], Aivaras Kareiva[4]

[1]Department of Applied Chemistry, Faculty of Chemistry, Vilnius University, Naugarduko 24, Vilnius LT-03225, Lithuania
[2]Institute of Materials Science, Faculty of Materials Science and Technology in Trnava, Slovak University of Technology in Bratislava, Paulinska 16, 91724 Trnava, Slovakia
[3]Laser Research Center, Department of Quantum Electronics, Physics Faculty, Vilnius University, Sauletekio Ave. 10, LT–10223 Vilnius, Lithuania
[4]Department of Inorganic Chemistry, Faculty of Chemistry, Vilnius University, Naugarduko 24, Vilnius LT-03225, Lithuania

## Abstract

The fabrication of polylactic acid (PLA)-carbonated hydroxyapatite (cHAP) composite material from synthesised phase pure nano-cHAP and melted PLA by mechanical mixing at 220-235 °C has been developed in this study. Topographical structuring of PLA-cHAP composite surfaces was performed by direct laser writing (DLW). Microstructured surfaces and the apatite distribution within the composite and formed grooves were evaluated by optical and scanning electron microscopies. The influence of the dopant concentration as well as the laser power and translation velocity on the composite surface morphology is discussed. The synthesis of carbonated hydroxyapatite (cHAP) nanocrystalline powders via wet chemistry approach from calcium acetate and diammonium hydrogen phosphate precursors together with crosslinking and complexing agents of polyethylene glycol, poly(vinyl alcohol) and triethanolamine is also reported. Thermal decomposition of the gels and formation of nanocrystalline cHAP were evaluated by thermal analysis, mass spectrometry and dilatometry measurements. The effect of organic additives on the formation and morphological features of cHAP was investigated. The phase purity and crystallinity of the carbonated apatite powders were evaluated by X-ray diffraction analysis and FT-IR spectroscopy.

# 1. Introduction

Bioactive ceramics, as an alternative to autografts (bones obtained from another anatomic site in the same subject) and allografts (bones from another subject of the same species), is a family of materials including various calcium phosphates, calcium carbonate, calcium sulphate and bioactive glasses. Such chemically active ceramics have the ability to bond to bones and enhance cell function and bone tissue formation within ceramic material. Carbonated hydroxyapatite (cHAP) – apatitic calcium phosphate – is a main inorganic constituent of bone and teeth tissue (50 to 70% apatitic mineral, 20 to 40% organic matrix, 5 to 10% water, and 3% lipids). [1] The carbonate groups can substitute both the hydroxyl ($OH^-$) and the phosphate ($PO_4^{3-}$) ions, resulting to the A-type ($Ca_{10}(PO_4)_6(OH)_{2-2x}(CO_3)_x$, where $0 \leq x \leq 1$) and B-type ($Ca_{10-y}(PO_4)_{6-y}(CO_3)_x(OH)_{2-x}$, where $0 \leq y \leq 2$) carbonation, respectively. [2-4] The presence of carbonate ions in the apatite crystal lattice not only defines the properties of produced material, but also determines the biological responses. For instance, it has been reported that as the carbonate content is increased, the sintering temperature required for a given degree of densification is decreased. [5] Studies have also showed that the dissolution rates of synthetic cHAP powders were higher in comparison to pure HAP and the bioactivity of samples increased with increasing carbonate content. [4] The higher osteoconductive properties and earlier bioresorption of the cHAP implants, as compared to HAP samples, were also reported. [6]

Conventionally, synthetic apatites are produced by a wet synthesis approach when precipitates of calcium phosphates in the amorphous or crystalline phase are directly formed under mixing calcium and phosphate ions. [7, 8] It has been demonstrated that many synthesis parameters like concentration, additives, reaction time, temperature, pH and stirring time affect the properties of the final product.[9-13] For example, the nature of precursor and organic additive may significantly alter the interconnection of inorganic species, formation of nucleus in the solution and subsequently affect the agglomeration of the formed particles. Therefore, the hydrophilic polymers (polyethylene glycol (PEG) or poly(vinyl alcohol) (PVA)) are often used in water-compatible polymer systems because of their proven ability to stabilize cations by multiple chelation. [9, 10] Recent studies showed that the content of the carboxy groups present in the nanofibers of natural hydrophilic polymer cellulose determine the heterogeneous nucleation of hydroxyapatite (HAP). [11] Phase purity and crystal shape can also be controlled by adding organic additives or by varying the pH of the solution. Han *et al*. demonstrated that increasing the concentration of the urea the apatitic phase transformation (dicalcium phosphate anhydrate ($CaHPO_4$, DCPA) – octacalcium phosphate ($Ca_8H_2(PO_4)_6 \cdot 5H_2O$, OCP) – HAP) can be obtained. [12]

The crystallinity of the material is another important aspect, since it governs dissolution properties and subsequently effects the biodegradation/bioresorption. [14] Temperature and pH are two major parameters that determine the crystallinity of ceramic material produced by precipitation method. There has been reported that precipitates formed from solutions at temperatures between 80 °C and 100 °C, the higher the initial pH, have a smaller fraction of calcium deficient material. Precipitation formed at temperatures < 80 °C leads to the less and less crystallized apatites. [15]

The degree of crystallinity can also be varied by changing the post-synthetic annealing temperature. [2, 16] Generally, the higher the annealing temperature is, the higher degree of crystallinity is observed. Studies performed by Juang and Hon demonstrated a poor resorbability of sintered hydroxyapatite-based materials and that was attributed to the loss of nanocrystallinity. [17]

Despite many advantages, apatites produced in the bulk form possess a low mechanical strength when used as a single constituent for load-bearing sites. Therefore, various bioceramic-polymer composites for bone-grafting material were prepared and relationship between material structure and properties has been studied. [18-20] Among synthetic polymers used in bone tissue engineering the polylactic acid (PLA) is probably the most widely investigated biodegradable and bioresorbable polyester. [21]

The degradation of bioceramic-polymer composite in a biological environment is a complex process and often is surface determined. So the surface modification of biomaterials has been a subject of many studies. Over the years different technologies have been used for topographical structuring to alter the roughness, surface energy/charge/area, and enhance the cells adhesion, morphology, differentiation, proliferation and osteoconductivity. [18, 22-27] For example, Zhang *et al*. demonstrated an effect of microgrooved polycaprolactone substrates, as compared to the flat substrates, on the cell migration along the microtrenches, which could in turn enhance the rate of bone fracture healing [28]. Similarly, De Luca *et al*. reported that microgrooved surfaces induce osteoblast maturation and protein adsorption [29]. Other studies have shown that adhesion and mobility of cells are governed by topographical factors such as height, slope and sharpness of the corners of microchannel walls [30].

Surface of biomaterials can be altered both chemically and physically. In recent decades among physical methods direct laser writing (DLW) gained a lot of attention due to its capability to be used as a versatile tool producing topographical micro- and nanostructures [25, 31, 32].

In this work cHAP nanocrystalline powders were prepared by wet chemistry method from calcium acetate and ammonium hydrogen phosphate precursors together with crosslinking agent of polyethylene glycol (PEG), poly(vinyl alcohol) (PVA) and triethanolamine (TEA). The decomposition of gels, apatite sintering temperature, its phase purity, crystallinity and chemical composition were examined by thermal analyses, X-ray diffraction analysis and FTIR spectroscopy. The effect of the chelating agent on the cHAP phase purity, particle and crystallite sizes is addressed. Further, PLA-cHAP composite material was prepared by incorporating synthesised nanocrystalline cHAP annealed at 680 °C into the PLA matrix by mechanical mixing at 220-235 °C. Topographical structuring of PLA-cHAP composite surfaces was performed by direct laser writing (DLW). The microstructured surfaces and the apatite distribution within the composite and formed grooves were evaluated by optical and electron microscopy coupled with energy-dispersive X-ray spectroscopy. The influence of the dopant concentration as well as the laser power and translation velocity on the composite morphology was investigated. The prepared PLA-cHAP composite could be a potential material for bone scaffold engineering.

## 2. Experimental

### 2.1. Synthesis of HAP powders

Calcium acetate hydrate ($Ca(CH_3COO)_2 \cdot \times H_2O$, ≥99 %, Roth) (5.285 g, 0.03 mol) and diammonium hydrogen phosphate (($NH_4)_2HPO_4$, 99.9 %, Alfa Aesar) (2.377 g, 0.018 mol) were first dissolved in separate beakers in 25 mL of distilled water. To these solutions 25 mL of 6% polyethylene glycol (H(OCH$_2$CH$_2$)$_n$OH, PEG, 4600; Aldrich) aqueous solution was then added as a crosslinking agent and the obtained mixtures were stirred for 30 min at 55-60 °C. In the following step ammonia aqueous solution (NH$_3$ (aq.), 32% aq. sol., Merck), was added to obtain pH = 11. These mixtures were additionally stirred for 20 min at 65-80 °C and then mixed together (Ca:P ratio of 1.67). The formed white Ca-P-O precipitate was further stirred for 30 min in a beaker covered with watch glass and latex cover to prevent ammonia evaporation. The obtained suspension was then evaporated at 80-100 °C and turned into Ca-P-O gel. The gel was dried at 150 °C for 24 h, calcined at 400 °C (heating rate 1 °C/min) and finally sintered at 600, 680, 800 and 1000 °C (heating rate 5 °C/min) for 5 h in air with intermediate grinding between each annealing temperature.

The same procedure was employed to prepare samples using crosslinking agents of poly(vinyl alcohol) ([-CH$_2$CHOH-]$_n$, PVA, 72000; Fluka) or PEG aqueous solution mixed with 1 mL triethanolamine ((HOCH$_2$CH$_2$)$_3$N, TEA, puriss. p.a., ≥ 99 % (GC); Sigma-Aldrich).

## 2.2. Preparation of PLA-cHAP composite

cHAP (PEG) powders sintered at 680 °C and polylactic acid (PLA, natural, 1.75 mm, DR3D Filament Ltd.) were used to prepare PLA-cHAP composite material. PLA was used as received from manufacturer/additional data not provided. cHAP powders of 5, 10, 15 and 20 wt% of the initial PLA weight were mixed mechanically with PLA in a beaker on hot plate at 220-235 °C. The obtained PLA-cHAP composites and neat PLA were then placed in a ceramic tile with 12 depressions (25 mm diam.) and additionally melted in furnace at 225 °C for 30 min (heating rate 300 °C/h) to obtain smooth pellets.

## 2.3. Topographical structuring of PLA-cHAP composite using DLW

The micro-cutting was performed using Yb:KGW laser "Pharos" (Light Conversion Ltd.) operating at 300 fs pulse duration, 200 kHz repetition rate and 515 nm wavelength (second harmonic). A high precision sample positioning and translation was realized using Aerotech linear stages. A 20x 0.45 NA objective lens was used for the beam focusing. More details on the employed setup can be found elsewhere [33]. The average laser power was measured as a main laser radiation parameter.

A schematic representation of laser-cutting algorithm is given in supplementary Fig. 1. The sample was translated in relation to the focal point in horizontal direction. The focal point was lowered towards the sample after every translation in horizontal plane. Such algorithm allowed to produce lines in the sample even if it's surface was not entirely even. The lines were fabricated using different translation velocities (250 µm/s, 500 µm/s, 750 µm/s and 1000 µm/s) and average laser powers of 5 mW, 10 mW, 15 mW and 20 mW. Laser power, $P$, was recalculated to energy per single pulse, $E$, and peak intensity at the focal point, $I$:

$$E = \frac{PT}{f}, \tag{1}$$

$$I = \frac{2PT}{fw^2\pi\tau} \tag{2}$$

In these formulas $P$ is the average laser power, $T$ is the objective transparency (0.41 for 20x 0.45 NA objective), $f$ is the pulse repetition rate, $\tau$ is the pulse duration, and $w = 0.61\lambda/NA$ is the waist (radius) of the beam.

Using these formulas we determined $P$, $E$ and $I$ used in this work (Table 1).

**Table 1.** Recalculation of measured average power $P$ to pulse energy $E$ and peak intensity $I$.

| $P$ [mW] | $E$ [nJ] | $I$ [TW/cm$^2$] |
|---|---|---|
| 5 | 10.25 | 4.47 |
| 10 | 20.5 | 8.93 |
| 15 | 30.75 | 13.40 |
| 20 | 41 | 17.86 |

As for translation characterization we estimated how many laser pulses reached one laser spot. To compute this we used the following formula:

$$n = \frac{2wf}{v}, \tag{3}$$

Where *v* is translation velocity. The number of laser pulses per spot is presented in Table 2.

**Table 2.** Translation velocity *v* and corresponding pulses per laser spot *n* used in this work.

| $v$ [μm/s] | $n$ |
|---|---|
| 250 | 1117 |
| 500 | 558.5 |
| 750 | 372.3 |
| 1000 | 279.2 |

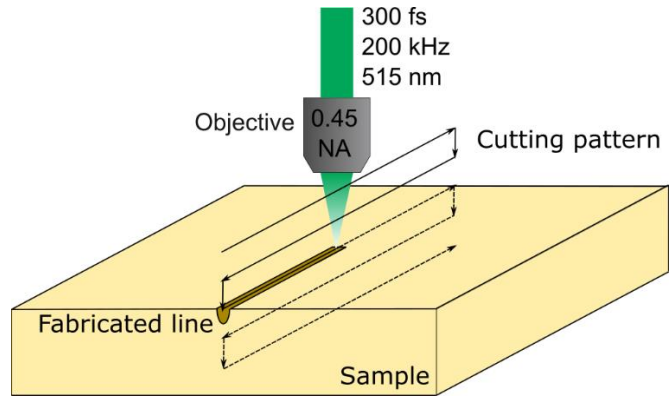

**Fig. 1**. Schematic representation of cutting pattern.

## 2.4. Characterization

Calcination temperature was evaluated by performing simultaneous thermogravimetry (TG), differential thermal analysis (DTA), derivative thermogravimetry (DTG) and mass spectroscopy (MS) measurements of the gels. A calibrated Netzsch STA 409CD instrument was employed. The gels were placed in alumina crucibles and heated from room temperature to 1300 °C with heating rate of $10°C.min^{-1}$. The sample chamber was evacuated before each measurement and purged with pure He (99.999 vol%). Then a stable gas flow of He (100 $ml.min^{-1}$) was set. The dimensional changes and phase transformations were studied by dilatometry analysis with Netzsch DIL 402C linear dilatometer. Isostatically pressed pellets of Ca-P-O gel powders in form of small cylinders were heated from room temperature to 1200 °C with heating rate of 3 °C.min$^{-1}$. Subsequently the samples were cooled to 100 °C by using a cooling rate of 10 °C.min$^{-1}$. The sample chamber was evacuated before each measurement and purged with pure Ar (99.9999 vol%). Then a stable gas flow of Ar (60 ml.min$^{-1}$) was set. The dilatometer was calibrated by an inert $Al_2O_3$ standard, which was measured at the same conditions as samples.

The phase of annealed powders was studied by X-ray diffraction (XRD, Rigaku, MiniFlex II, Cu-Kα radiation, λ = 0.1542 nm, 40 kV, 100 mA, 2$\Theta$ = 10-80°/10-60°). The average crystallite sizes were determined by Halder-Wagner method in the Rigaku PDXL software. Morphological features and elemental distribution were evaluated using field emission scanning electron microscopy (FE-SEM, SU70, Hitachi, 5.0 kV acc. voltage) and tabletop scanning electron microscope (SEM, TM3000, Hitachi, 15.0 kV acc. voltage) equipped with the energy dispersive X-ray spectrometer (EDX). X-ray accusation times of 60 and 122 seconds were used to obtain the EDX spectra and elemental mapping images (256 × 192 pixels, process time 5), respectively. To examine the morphology of the formed grooves, PLA-cHAP composites were coated with gold (sputtering time 30 s) using sputter coater (Q15OT ES, Quorum). Infrared spectra were recorded using Fourier transform infrared (FT-IR) spectrometer (Frontier FT-IR, Perkin-Elmer) equipped with Gladi attenuated total reflection (ATR) viewing plate (Diamond ATR crystal) and Liquid-nitrogen-cooled mercury cadmium telluride (MCT) detector (4000-500 cm$^{-1}$, 25 scans). Samples were also characterized via confocal microscope/profilometer PLµ 2300 (Sensofar) using 20x 0.45 NA objective.

## 3. Results and discussion

### 3.1. Evaluation of sintering temperature

The sintering temperature of cHAP was evaluated from TG and DTA analyses of Ca-P-O gel powders. TG and DTA curves of the representative Ca-P-O gel powders synthesised using PEG are presented in Fig. 2(a). The first very small weight loss of ~ 0.5% was observed at temperatures up to 200 ºC and was assigned to the removal of adsorbed water. The second small weight loss of 2% was observed up to 400 ºC and was assigned to the absorbed water and partial decomposition of organic compounds. Some lattice water and $CO_2$ were released in the temperatures 400 to 600 °C. Such processes could be seen in the DTG curve where broad peak with a maximum at ~ 550 °C is obtained (~1% weight loss). Next weight loss of ~ 3% is observed at 600 to 900 ºC and simultaneously broad exothermic reaction (DTG curve) with a maximum at 710 ºC takes places. This could be attributed to the further lattice dehydration and decarboxylation as well as crystallization of cHAP. It has to be noted that the crystallization of cHAP occurs over a range of temperatures and is dependent on the solution composition and processing conditions. The phase transformation of amorphous phase to crystalline hydroxyapatite was reported by K. Gross [34]. Furthermore, previous studies showed that in the temperature range from 600 °C to 1000 °C cHAP loses a major part of its carbonate [35]. Other weight loss observed between 1150-1250 ºC is less significant, but small exothermic peaks in the DTA curve (maxima at 1180 and 1280 ºC) are initiated due to decomposition of cHAP and possible formation of CaO and other secondary phases in the material. A similar thermal behaviour was obtained of Ca-P-O gels synthesised using PEG-TEA and PVA matrices. Thermogravimetric analysis, as shown in Fig. 3, demonstrates the processes such as adsorbed water and volatile organic compound loss up to 400 °C, lattice dehydration and decarboxylation (broad exothermic peaks ~ 700 °C) which indicates a possible reorganization in the structure.

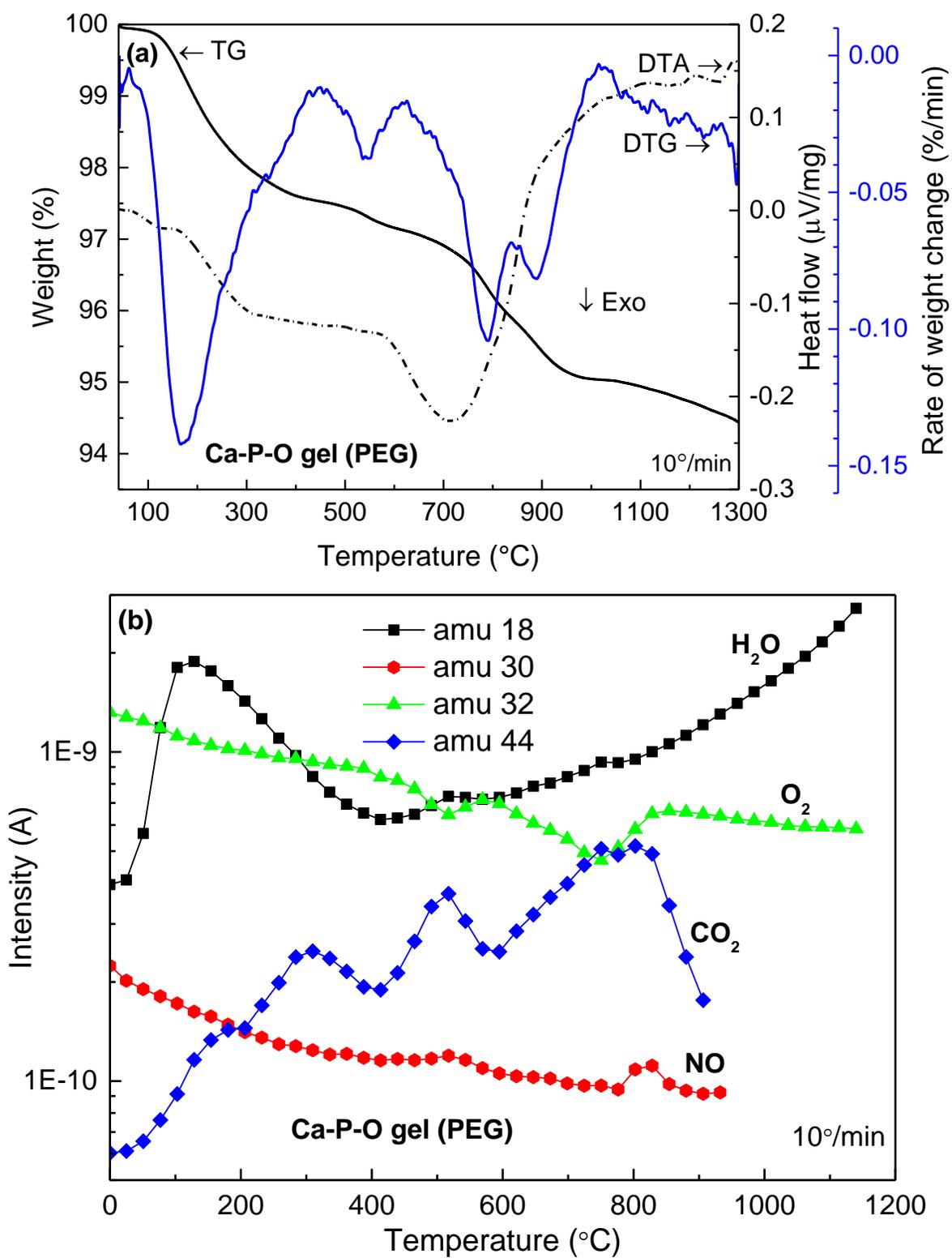

Fig. 2. (a) TG-DTG, DTA and (b) MS curves of Ca-P-O (PEG) gel.

MS fragment ion curves (Fig. 2(b)) recorded of the Ca-P-O gel (PEG) powders for $H_2O$ (m/z = 18), NO (m/z = 30), $O_2$ (m/z = 32) and $CO_2$ (m/z = 44) show peaks that correspond to the individual TGA steps. The major release of adsorbed $H_2O$ in the MS curve was observed below 200 °C with a less prominent dehydration peaks appearing ~ 520 and 780 °C. The $CO_2$ elimination from the sample was observed parallel to dehydration. Studies showed that the processes appearing below 850 °C are irreversible decomposition of material. Both $CO_2$ and $H_2O$ elimination upon heating gives carbonate relocation and subsequent formation of CaO and B-type HAP transformation to the A-type [35]. The evolution process of $H_2O$ and $CO_2$ in carbonated HAP was demonstrated and discussed by Yasukawa *et al.* where desorption of both $H_2O$ and $CO_2$ adsorbed on the particles below 600 °C and liberation of these molecules above 700 °C occur. [36]. MS curves of Ca-O-P (PEG-TEA) and Ca-P-O (PVA) gel powders produced in this work showed similar features (Supplementary Fig. S1 and Fig S2).

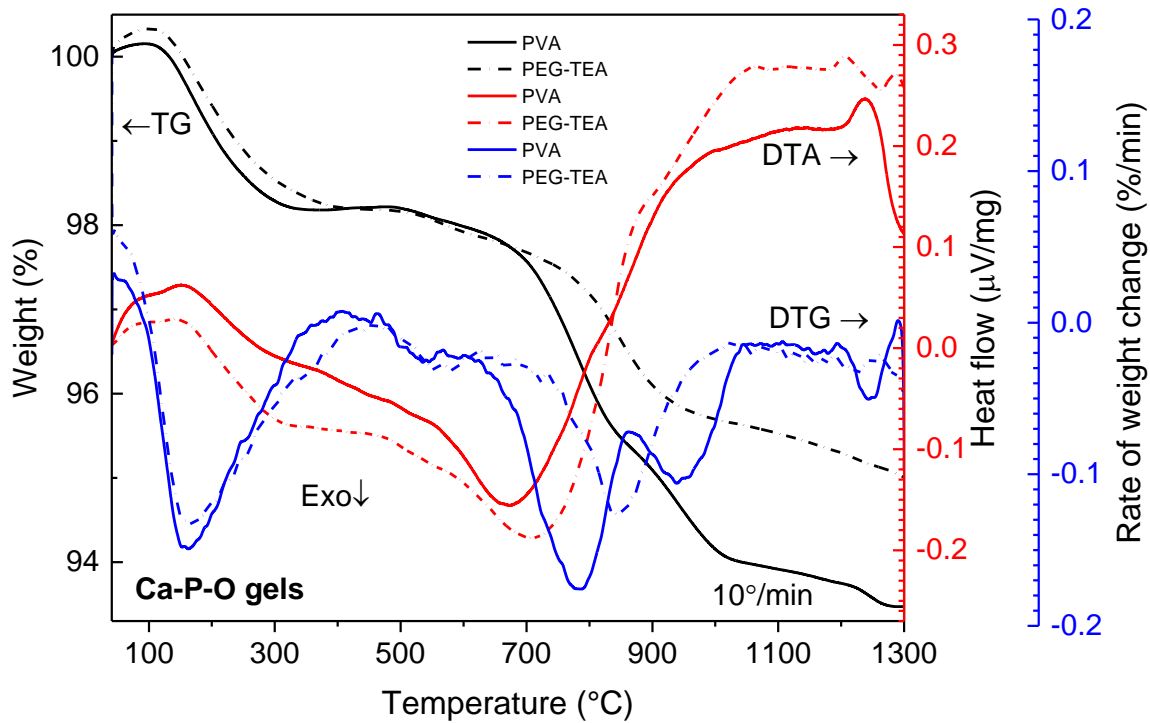

**Fig. 3.** TG-DTG and DTA curves of Ca-P-O gel powders synthesized using PVA and PEG-TEA matrices (at 10°/min up to 1300 °C).

The calcination temperature and structural changes of cHAP powders were further studied by dilatometry (Fig. 4). Dilatometry results show that relative length change *dl/l* of such materials vary with temperature and polymer matrix used in the synthesis process.

The variation of the length L in the pellets consist of two parts for the powders produced using PEG and of three parts for the powders produced using TEA-PEG and PVA polymer matrix. The first contraction from 100 to 200 °C corresponds to the initial elimination of volatile compounds in agreement with TG and DTA results. The second part in the Ca-P-O samples is connected to a second contraction and corresponds to the decomposition and elimination of absorbed water and organics, as well as transformation of ions and crystallization of HAP. Results show that temperature region for the second contraction is significantly influenced by the polymer matrix used in the synthesis process. Dilatometry curve of Ca-P-O (PEG) samples exhibit a linear shrinkage from 500 to 1100 °C, while for the Ca-P-O (PEG-TEA) and Ca-P-O (PVA) samples a shrinkage was observed in the temperature regions of 500-850 °C and 480-800 °C, respectively. The third contraction for the Ca-P-O (PEG-TEA) and Ca-P-O (PVA) samples was observed in the temperature regions of 900-1200 °C and 800-1200 °C, respectively, and could be assigned to the decomposition of HAP and formation of CaO and other secondary phases in the material. It can also be seen that at the dwell temperature of 1100 °C there is no shrinkage, indicating that the densification process of all samples at this temperature is terminated. Therefore, in order to prevent the carbonate loss from the material as well as formation of CaO, the produced gels were calcined at 400-800 °C.

One can also observe that shrinkage of Ca-P-O (PVA) begins at lower temperatures in comparison to Ca-P-O (PEG) or Ca-P-O (TEA-PEG) (Fig. 4). This observation indicates that volume or grain boundary diffusion appears at lower temperatures for material derived from PVA matrix. This further indicates that the material produced PEG matrix at the same temperatures might possess looser particles. A similar investigation showing an effect of calcium phosphate nature on the sintering rate was reported by Champion. [37]

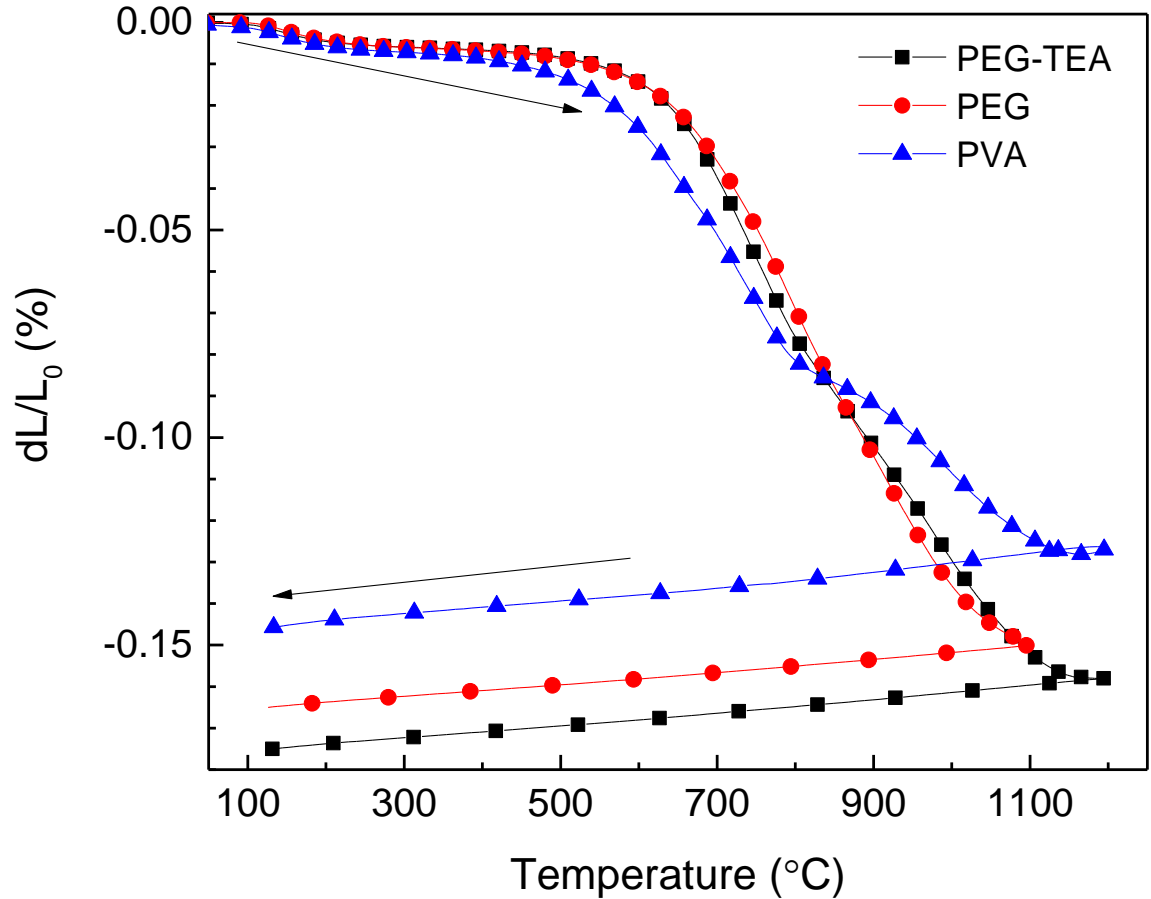

**Fig. 4.** Thermal dilatometry curves of the cHAP powders (synthesized using PEG, PEG-TEA and PVA): linear shrinkage *vs* temperature.

### 3.2. Crystallinity and phase purity of bulk samples

XRD patterns of Ca-P-O (PEG) gel powders annealed at different temperatures are presented in Fig. 5(a). It can be observed that powders produced at 600 °C are already crystalline and the intensity of the diffractions increased with increasing annealing temperature (800 °C). The sharply resolved diffraction peaks arise from the Bragg diffraction conditions for HAP (JCPDS no. 09-0432, hexagonal crystal system, space group of P6$_3$/*m*) which is consistent with literature (data not presented). [2] The crystallite size for powders annealed at 600, 680 and 800 °C was calculated to be 17, 22 and 26 ± 1 nm, respectively. The unit cell parameters a = b and c of HAP (PEG) were calculated to be 9.439 Å and 6.898 Å, respectively.

Additionally, to estimate the Ca/P ratio of apatite the synthesised material was annealed at 1000 °C. It is widely known that nonstoichiometric apatites are not stable and decompose at 900-1100 °C producing stoichiometric HAP and additional phases such as CaO and tricalcium phosphate (TCP). It is established that Ca/P ratio is higher than 1.67 (stoichiometric HAP), if heated sample is composed of pure HAP and CaO. The diffraction pattern of HAP (PEG) sample annealed at 1000 °C (Supplementary Fig. S3) showed that powders contain HAP and CaO suggesting that nonstoichiometric HAP was formed, with Ca/P ratio larger than 1.67.

The XRD patterns of Ca-P-O (PEG-TEA) and Ca-P-O (PVA) gel powders annealed at 800 °C (Fig. 5(b)) show peaks of crystalline HAP. A noticeable diffraction peak at $2\theta = 37.4°$ was also observed in the diffraction patterns of powders annealed at 680 °C (data not presented) and 800 °C and was assigned to the impurity phase of CaO. The background increment obtained in the region from 15 to 30 $\theta$ degree (XRD pattern of the powders synthesized using PEG-TEA) is due to the glass sample holder. The average crystallite sizes for the HAP powders synthesized using PVA and PEG-TEA were estimated to be 43 and 39 ± 1 nm, respectively.

The obtained XRD results indicate that phase composition and crystallite sizes are affected by the polymer nature, its initial concentration, solution pH and suspension mixing time. Studies carried out by A. L. Boskey and A. S. Posner showed that the transformation of amorphous calcium phosphate to hydroxyapatite is solution mediated and depends on the conditions which regulate both the dissolution of amorphous calcium phosphate and the formation of the preliminary hydroxyapatite nuclei. [38]

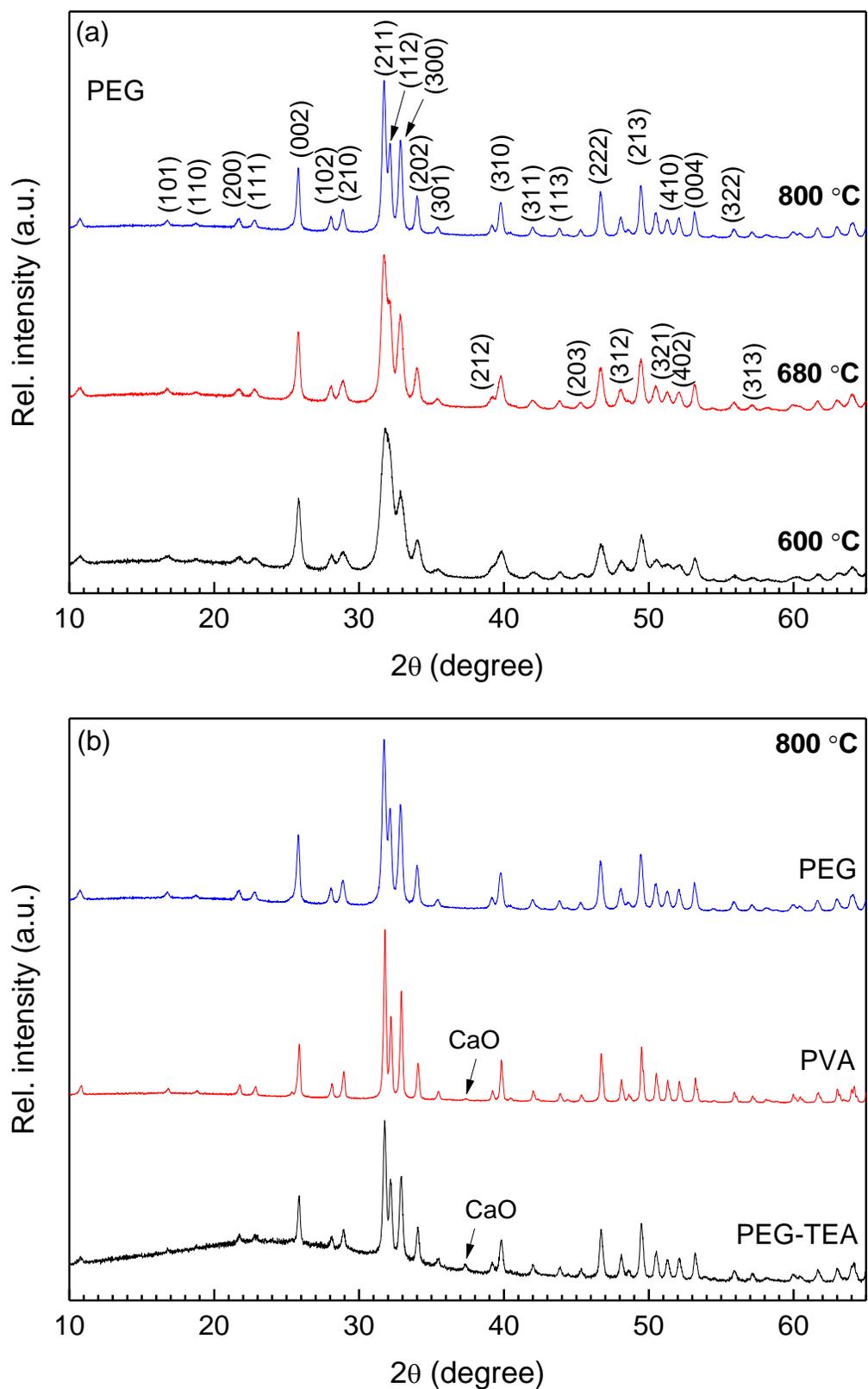

**Fig. 5.** XRD diffraction patterns of HAP powders (a) synthesized using PEG and calcined at 600, 680 and 800 °C and (b) synthesized using PEG-TEA, PVA and PEG (calcined at 800 °C) showing impurity phase of CaO (background for HAP (PEG-TEA) powders is elevated due to the glass nature of sample holder).

### 3.3. Structure and carbonate determination through IR spectroscopy

FT-IR spectra of apatite synthesized using PEG show characteristic bands of the phosphate ($PO_4^{3-}$) and carbonate ($CO_3^{2-}$) groups (Fig. 6). [2, 39] The absorption bands in the regions of 1190-900 cm$^{-1}$ are due to the triply degenerated asymmetric stretching mode, $v_3$, and symmetric stretching mode, $v_1$, of the P–O bonds, while bands obtained in the region of 630-530 cm$^{-1}$ comes from triply degenerated bending mode, $v_4$, of the O–P–O bonds of apatitic $PO_4^{3-}$ groups. Bands in the 1550-1360 cm$^{-1}$ region are characteristic for the $CO_3^{2-}$ groups. A band located at the 1466 cm$^{-1}$ was assigned to the stretching modes, $v_1$, of $CO_3^{2-}$ group in A-type substituted cHAP. The characteristic bending modes, $v_4$ or $v_3$, of C–O bond ($CO_3^{2-}$ group) in A- and B-type cHAP appears at 1456 and 1445 cm$^{-1}$. The band at 1420 cm$^{-1}$, which is followed by another weak band at 1422 cm$^{-1}$ were assigned to the stretching modes, $v_3$, of the $CO_3^{2-}$ group in the cHAP. In addition, a weak band arising from the bending mode, $v_2$, of C–O bond ($CO_3^{2-}$ group) was observed at ~ 875-880 cm$^{-1}$. In the 530-630 cm$^{-1}$ region, bands located at ~ 599, 565 cm$^{-1}$ were assigned to the triply degenerated bending modes, $v_{4a}$, and $v_{4c}$, of the O−P−O bonds of the phosphate group. Furthermore, two distinctive peaks observed at 3573 and 630 cm$^{-1}$ assigned to the stretching mode, $v_s$, and vibrational mode, $v_L$, of the structural hydroxyl anion (OH$^-$) in cHAP, respectively. The broad absorption band at ~3500-3060 cm$^{-1}$ (maximum at 3242 cm$^{-1}$) appears due to the adsorbed water molecules. The assignment of all obtained bands in FT-IR spectra is summarised in Table 3. Apatites produced using PEG-TEA or PVA polymeric matrices exhibited similar spectral features (data not presented).

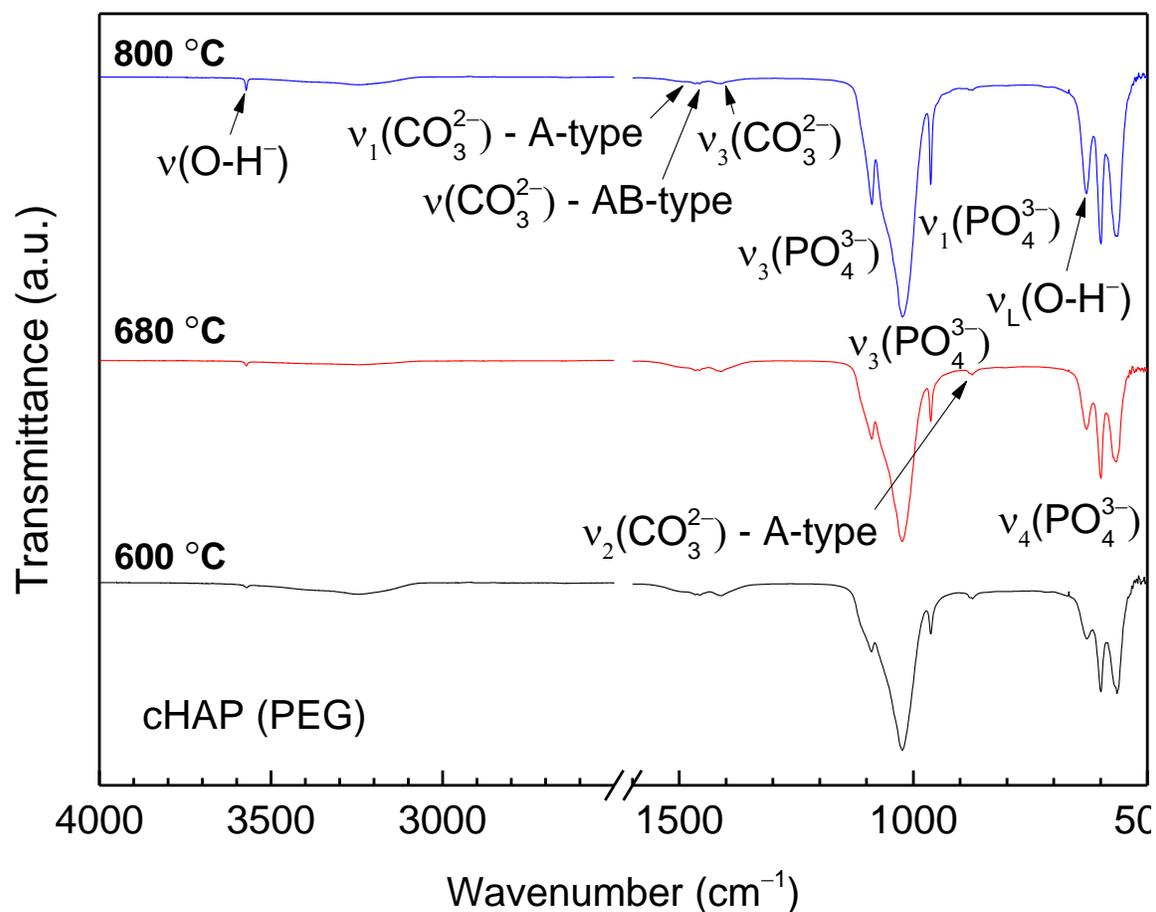

**Fig. 6.** FT-IR spectra of cHAP samples synthesized using PEG and calcined at 600, 680 and 800 °C.

**Table 3**. IR band assignment of the cHAP (PEG) annealed at 800 ºC.

| Wavenumber (cm$^{-1}$) | Peak assignment | Reference |
|---|---|---|
| 3573 (w) | $\nu_s$, stretching mode, of O–H | 2, 39, 40 |
| 1466 (m) | $\nu_1$, stretching modes, of $CO_3^{2-}$ (A-type) | 2, 39-42 |
| 1456 (vw/sh) | $\nu_3$ or $\nu_4$, bending mode, of $CO_3^{2-}$ (A and B type) | 2, 39, 41 |
| 1445 (s) | $\nu_3$ or $\nu_4$, bending mode, of $CO_3^{2-}$ (A and B type) | 2, 41-43 |
| 1420 (m) | $\nu_3$, stretching mode, of $CO_3^{2-}$ (B type) | 2, 39, 41-43 |
| 1411 (s) | $\nu_3$, stretching mode, of $CO_3^{2-}$ (B type) | 2, 39, 43 |
| 1088 (s) | $\nu_{3a}$, triply degenerate asymmetric stretching mode, of $PO_4^{3-}$ (P–O bond) | 2, 39 |

| 1031 (s) | $\nu_{3c}$, triply degenerate asymmetric stretching mode of $PO_4^{3-}$ (P–O bond) | 2, 39 |
| --- | --- | --- |
| 1023 (vs)/1029 (sh) | Due to the presence of crystal imperfections and $HPO_4^{2-}$ groups | 39, 44 |
| 963 (m) | $\nu_1$, nondegenerated symmetric stretching mode, of $PO_4^{3-}$ (P–O bond) | 39, 43 |
| 880 (m)/875 (w) | $\nu_2$, bending mode, of $CO_3^{2-}$ (A type) | 2, 39, 40, 42, 43 |
| 630 (m) | $\nu_L$, librational mode, of OH (O–H bond) | 2, 39, 40 |
| 599 (s) | $\nu_{4a}$, triply degenerate bending mode, of $PO_4^{3-}$ (O–P–O bond) | 2, 39, 43 |
| 565 (ws) | $\nu_{4c}$, triply degenerated bending mode, of the $PO_4^{3-}$ (O–P–O bond) | 2, 39, 43 |

**3.4 Morphology examination of cHAP**

FE-SEM micrographs of cHAP powders produced using PEG, PEG-TEA and PVA are presented in Fig. 7. Powders produced at 680 °C using PEG matrix are build up of agglomerates with primary particles of about 20-30 nm (Fig.7(a)). Further annealing at 800 °C resulted in particle growth. It is interesting to note, that a slight elongation of primary particles was observed of the powder samples produced using PEG matrix. Furthermore, estimated cHAP (PEG) particle sizes correspond well with crystallite sizes calculated from XRD data. This suggests that synthesized particles are single crystalline. Particles produced using PEG-TEA matrix exhibited similar morphological features (Fig. 7(c)). However, it has to be noted that the marginal increase in particle size and slightly larger neck formation between interconnected particles comparing to cHAP (PEG) has been observed. This correlates with TG data discussed above where slightly lower reaction temperature in the region of 700-800 °C was observed for the cHAP (PEG-TEA) powders. Completely different morphological features were obtained for powders derived from PVA matrix (Fig. 7(d)). One can see that agglomerated particles are compact and consist of primary particles larger than 100 nm. While particles produced from PEG and PEG-TEA matrix have mostly spherical shape, cHAP (PVA) particles possess more faceted particles. Results are in consistence with dilatometry data.

Furthermore, the Ca/P ratio estimated by EDX analysis varied with predominant value larger than 1.67. Studies showed that for nonstoichiometric apatites, when $PO_4^{3-}$ ions are replaced by $CO_3^{2-}$ ions (B-type substitution), Ca/P ratio is larger than that of stoichiometric HAP. However, when $OH^-$ ions in the apatite structure are replaced by $CO_3^{2-}$ ions (A- type substitution) or $HPO_4^{3-}$ ions replace $PO_4^{3-}$ the Ca/P ratio tend to be smaller than 1.67. [36, 44] EDX results support XRD and IR findings that A- and B-type substituted cHAP were produced.

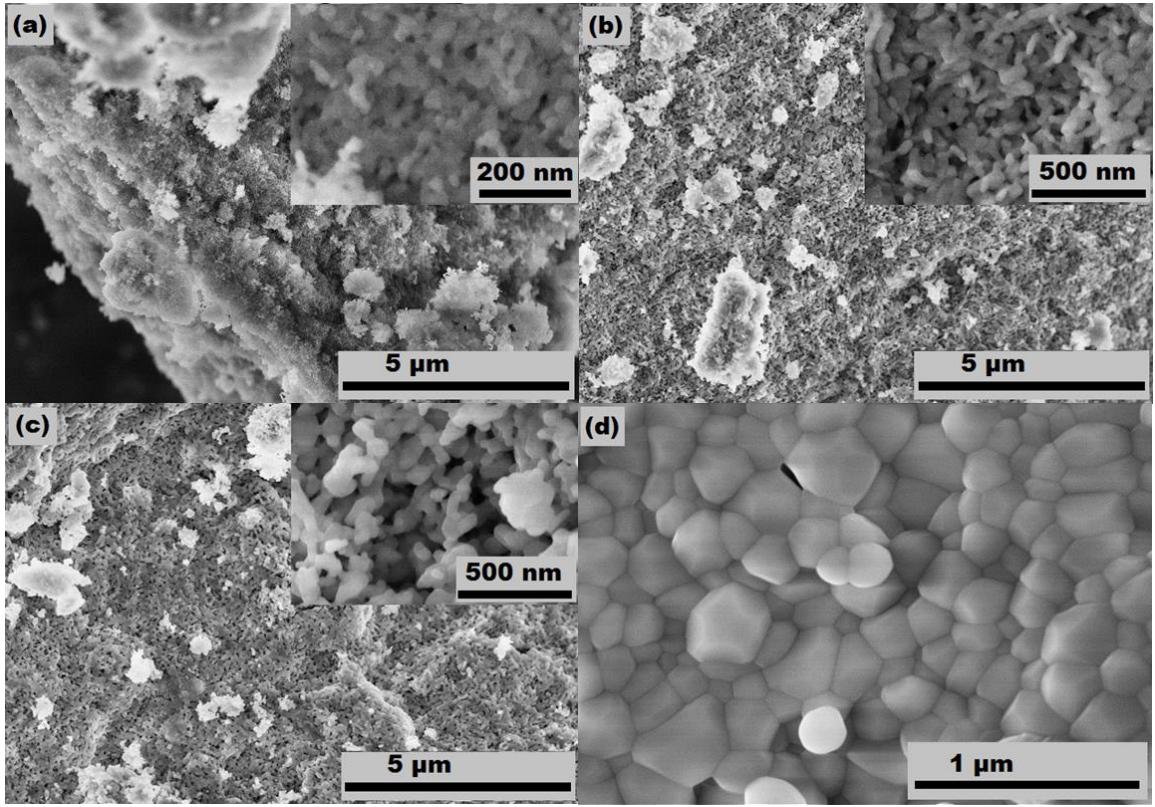

**Fig. 7.** FE-SEM micrographs of cHAP powders synthesized at (a) 680 °C and (b) 800 °C using PEG, and at 800 °C using (c) PEG-TEA and (d) PVA.

**3.5. PLA-cHAP composites topographical structuring using DLW**

Micrographs of lines fabricated with aforementioned algorithm (Fig. 8) on the neat PLA and PLA-cHAP surfaces led to several important observations. Firstly, the lines produced in neat PLA seem to have even sides (Fig.8(a)). This corresponds to the previous studies already reported in literature. [45, 46] Only their width changed when fabrication parameters (translation velocity or average laser power corresponding to pulse overlap and their energy/peak intensity, respectively) were varied. On the other hand, materials doped with cHAP showed a tendency to react to laser light in a more violent manner and started to melt uncontrollably. The melting itself showed to be inhomogeneous and started at random points of the cut. This indicates that such material has a higher and uneven sensitivity to the light, notably when 13.4 or 17.86 TW/cm$^2$ peak intensity was used for microfabrication (Fig. 8(f)). Finally, if radiation intensity was relatively low (4.47 and 8.93 TW/cm$^2$) (Fig. 8(a-e)) all cuts seemed to have even sides and only their widths were different.

The several possible reasons might be assumed for PLA-cHAP material sensitivity to the light. Previous studies have shown that there are double melting peaks in neat PLA [47, 48]. PLA melting and remelting peaks as well as crystallization temperature can vary according to PLA nature and inorganic filler introduced in the PLA matrix. For example, Shi and Dou compared the crystallization behaviour of neat PLA and PLA matrix filled with inorganic nanocrystalline $CaCO_3$ showing that crystallization temperature of PLA tend to be slightly reduced after the filling of inorganic material [47]. Similarly, studies carried out by Shakoor and Thomas showed that the cold crystallization temperature PLA/talk composite was reduced and crystallinity was increased after addition of talk to the PLA matrix [49]. This might be one of the reasons why PLA-cHAP material is sensitive to the light.

Another interesting point should be mentioned. In the current work the processing temperature was increased by several degrees to obtain homogeneous PLA-cHAP composite when 10 or 15 wt% of cHAP was added to the melted PLA. This might also be coursing some structural changes in the PLA matrix and subsequently resulting in a higher sensitivity to the light. Studies of the PLA-cHAP composite sensitivity to the light is, however, deemed to be outside the scope of this work, and will be the subject of further studies.

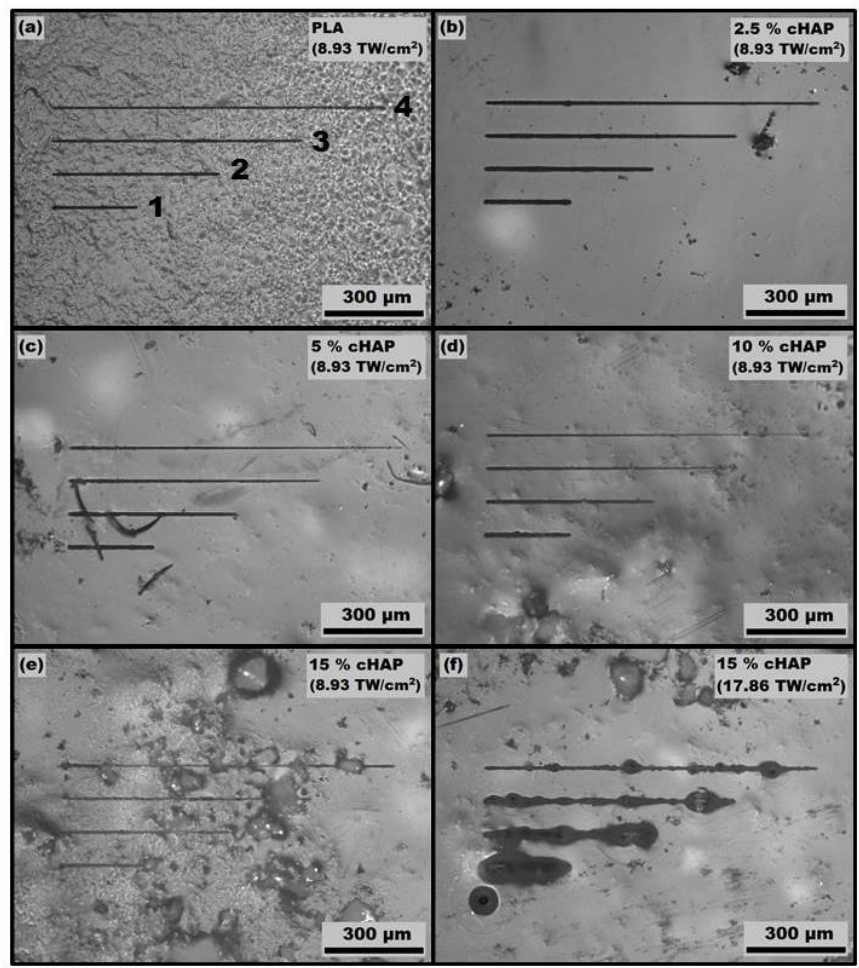

**Fig. 8.** Optical micrographs of lines fabricated on the surface of (a) PLA and PLA-HAP composite pellets. Pellets were prepared using (b) 2.5%, (c) 5%, (d) 10% and (e and f) 15% of HAP surface. Lines were fabricated using peak intensity of 8.93 TW/cm$^2$ for (a)-(e) samples and 17.86 TW/cm$^2$ for (f) sample (the translation velocities of 250, 500, 750 and 1000 µm/s were used to produce lines in increasing length order).

### 3.6. Surface morphology of femtosecond laser ablated PLA-cHAP composites

To obtain a better knowledge of cHAP distributtion within the formed composite we performed SEM and EDX analyses (Fig. 9). Morphological studies showed that with the increasing cHAP content the homogenious distribution of inorganic matter within PLA-cHAP composite decreased. Biocomposites containing a larger amount of cHAP exhibit more patches formed of inorganic matter distributed over the entire composite surface (Fig. 9(a)). The EDX elemental mapping was performed to estimate the Ca and P – cHAP – distribution on the composite surface and formed grooves (Fig. 9(b)-(i)). One can see, that the cHAP amount varies within different surface regions examined. Despite heterogeniuos distribution of inorganic matter formed within high percentage of doped biocomposites, the interesting observation is that Ca and P become dominant elements within the cuts. This indicates that cHAP particles are covered with a polymer layer and, after the ablation, the inorganic apatite matter becomes exposed to the surface. Such topografical structuring and observed features might be beneficial in terms of cell adhesion and profiliteration for scaffold engineering. A recent experimental study by Zhao *et al.* showed that micropatterned HAP surfaces promote the growth and osteogenic differentiation of bone marrow stromal cells [27].

Further, it shall be noted, that PLA-cHAP surface sites examined by SEM erode under the beam exposure (marked circles in Fig. 9(f) and (h)), leaving rough (lumpy) patches of melted PLA with diminished grooves. Analysis of different locations gave very similar results.

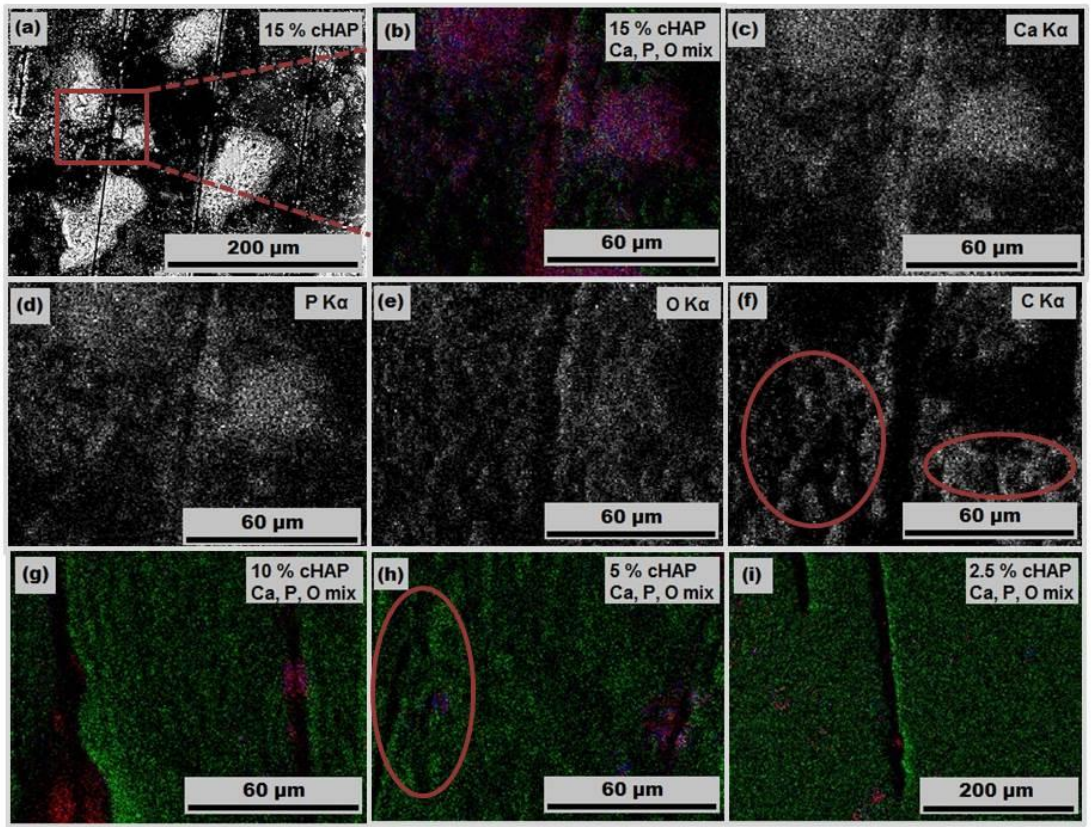

**Fig. 9**. (a) SEM (marked place shows the surface region examined) and (b-i) EDX elemental mapping micrographs of PLA-HAP composites doped with (a-f) 15%, (g) 10%, (h) 5% and (i) 2.5% of HAP (designated EDX mapping colours: Ca – red, P – blue and O

– green). The marked places in (f) and (h) micrographs show the damaged regions of PLA-cHAP composite surfaces.

FE-SEM micrographs of fabricated microtrenches are shown in Fig. 10 (prior to examination composite pellets were metalised). One can see that the peak intensity of 4.47 and 8.93 TW/cm$^2$ used for surface topographical structuring give narrower and tidier grooves (1, 2, 3 and 4 lines produced using 250, 500, 750 and 1000 µm/s, respectively) in comparison to those produced with peak intensity of 13.4 and 17.86 TW/cm$^2$ (Fig. 10(a)-(d)). The widths of the cuts were estimated to be ~ 20, 15, 10 and 5 µm and are coincident with the laser power in diminishing order used to produce lines. Grooved surfaces of neat PLA exhibited similar morphological features as those produced at a low doping concentration (2.5% cHAP) (data not presented). In the case of higher cHAP doping concentrations, as discussed previously, the overall surface of materials become rougher and possess domains (~ 50-150 µm) of cHAP exposed to the surface (Fig. 10((i), (o)). The quantity and diameter of cHAP domains increases with increasing dopant concentration. Micrographs of cuts formed on composites having higher amount of dopant show that the femtosecond laser ablation exposes cHAP nanoparticles to the surface while slightly larger cHAP crystals were obtained in some areas between grooved lines (Fig. 10(e)-(m), (p)).

Moreover, micrographs revealed that over entire 15 and 20 µm width produced lines more sections with deeper valleys and melted edges has been obtained in comparison to those of 5 and 10 µm (Fig. 10((a), (e), (f), (q)). Furthermore, it was observed that surface material melting under laser radiation appears in homogeneous areas and those where cHAP is visible (Fig. 10((a), (e), (f), (q)). The same feature was observed in all samples. Thus, the results suggest that surface material melting might be due to the air trapped inside the composite that formed voids within some areas of composite. Due to these voids material internal walls vary in thickness and, as a result, respond differently to the laser light. On the contrary, one will find that the formed voids and different thickness domains might lead to some porosity which is one of the most important characteristics for scaffolding materials.

Although we can conclude that in the current work the most evenly cut grooves with nanoparticles of cHAP homogeneously exposed to the surface were obtained using 4.47 and 8.93 TW/cm² peak intensity for 10 and 15% cHAP doping, the studies examining the direct effect of the observed surface topography on cell behaviour *in vitro* and *in vivo* still has to be performed and will be a subject of further studies.

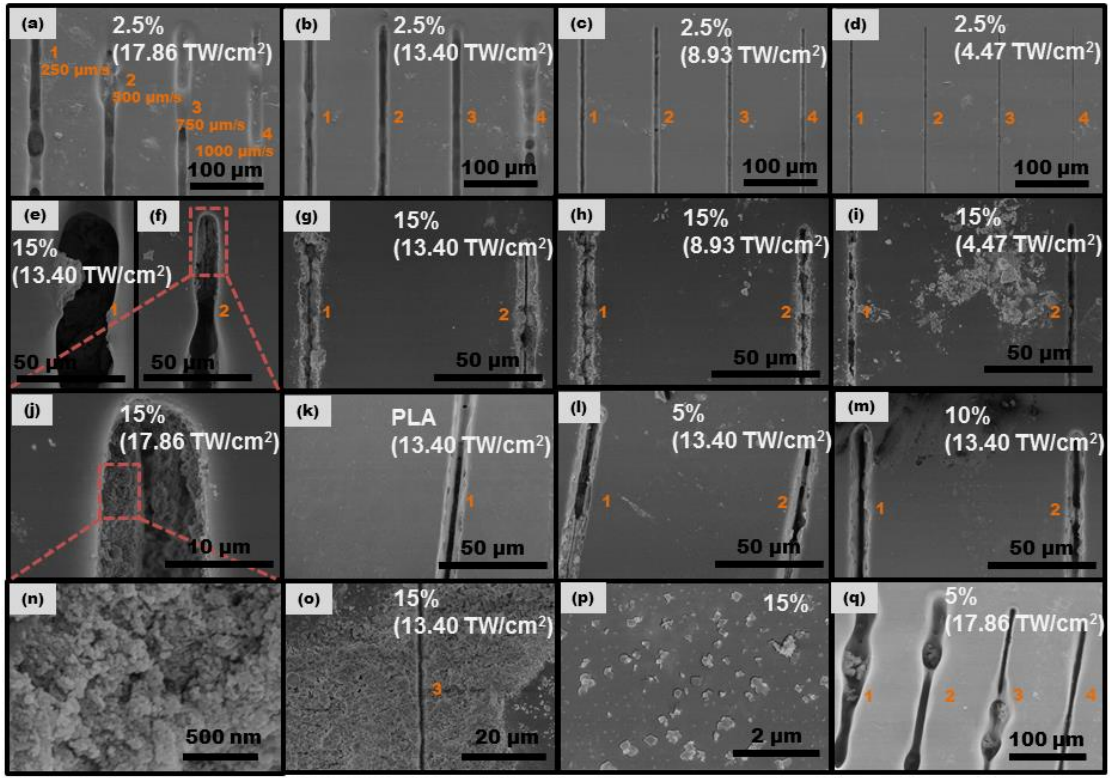

**Fig. 10.** FE-SEM micrographs of PLA-cHAP composite grooved surfaces: (a)-(d) 2.5% of cHAP; (e)-(j) 15% of cHAP; comparison of the grooves produced with the peak intensity of 13.4 TW/cm² of (k) neat PLA, (l) 5% cHAP and (m) 10% cHAP doping; (n) cHAP particles inside the formed groove (15% cHAP); (o) showing larger domain of cHAP exposed to the surface (15% of cHAP); (p) showing larger cHAP crystals obtained between grooved lines; (q) observed surface material melting under laser radiation (5% cHAP doping).

## Conclusions

Homogeneous nanocrystalline cHAP with tailored morphological and structural properties were produced by wet chemistry approach and varying organic additives. The particle growth was strongly influenced by the chelating agent used in the synthesis process. The results of thermal analysis revealed that the densification behaviour of synthesised powders depends on the produced material particle/grain sizes and phase composition. XRD results showed that powders calcined at 680 °C are crystalline, while annealing at higher temperatures induced particle growth and rise of CaO phase. The investigations of Ca/P ratio showed that nonstoichiometric apatite is formed at 680 °C. FT-IR analysis revealed the formation of A- and B-substituted cHAP. From FE-SEM analysis it was evident that the most evenly cut grooves with nanoparticles of cHAP homogeneously exposed to the surface for 10 and 15wt.% cHAP-PLA composites were obtained using higher translation velocities (750 μm/s and 1000 μm/s) and lower laser radiation intensities (4.47 TW/cm$^2$ and 8.93 TW/cm$^2$). EDX analysis showed a homogeneous distribution of cHAP within the cut. The prepared PLA-cHAP composite could be a potential material with applications for bone scaffold engineering.

**Acknowledgement**

EG acknowledge the SAIA for provided scholarship under the National Scholarship Programme of the Slovak Republic for the Support of Mobility of Students, PhD Students, University Teachers, Researchers and Artists. This research was partly funded by a grant (No. MIP-046/2015) from the Research Council of Lithuania.

**Supporting information - figure captions:**

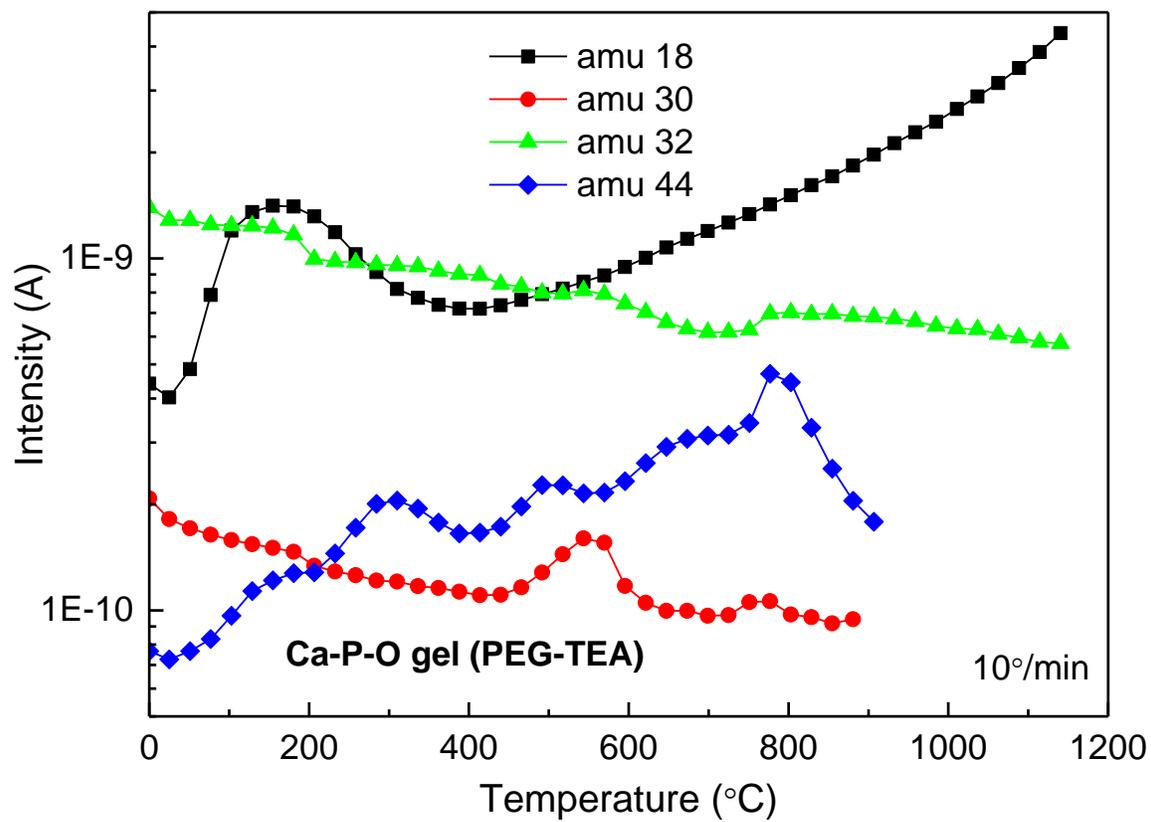

**Figure S1.** MS curves of Ca-P-O gel synthesized using PEG-TEA matrix.

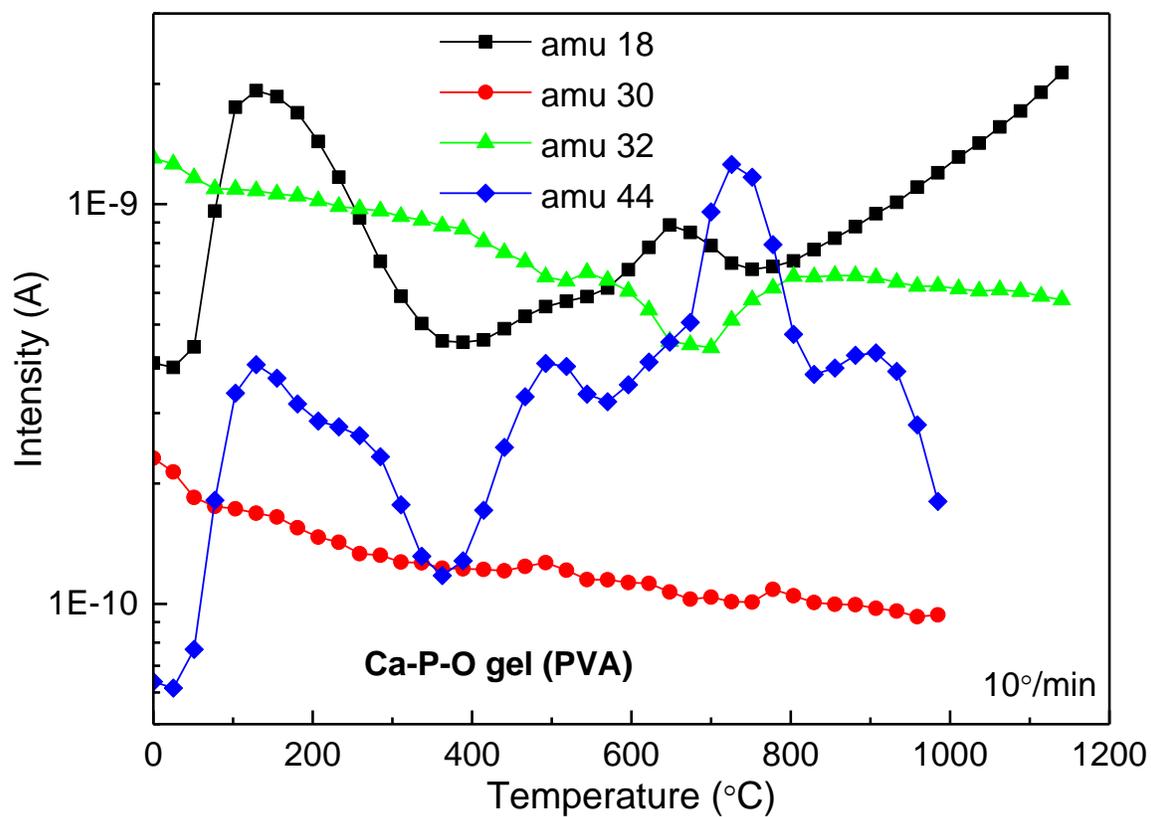

**Figure S2.** MS curves of Ca-P-O gel synthesized using PVA matrix.

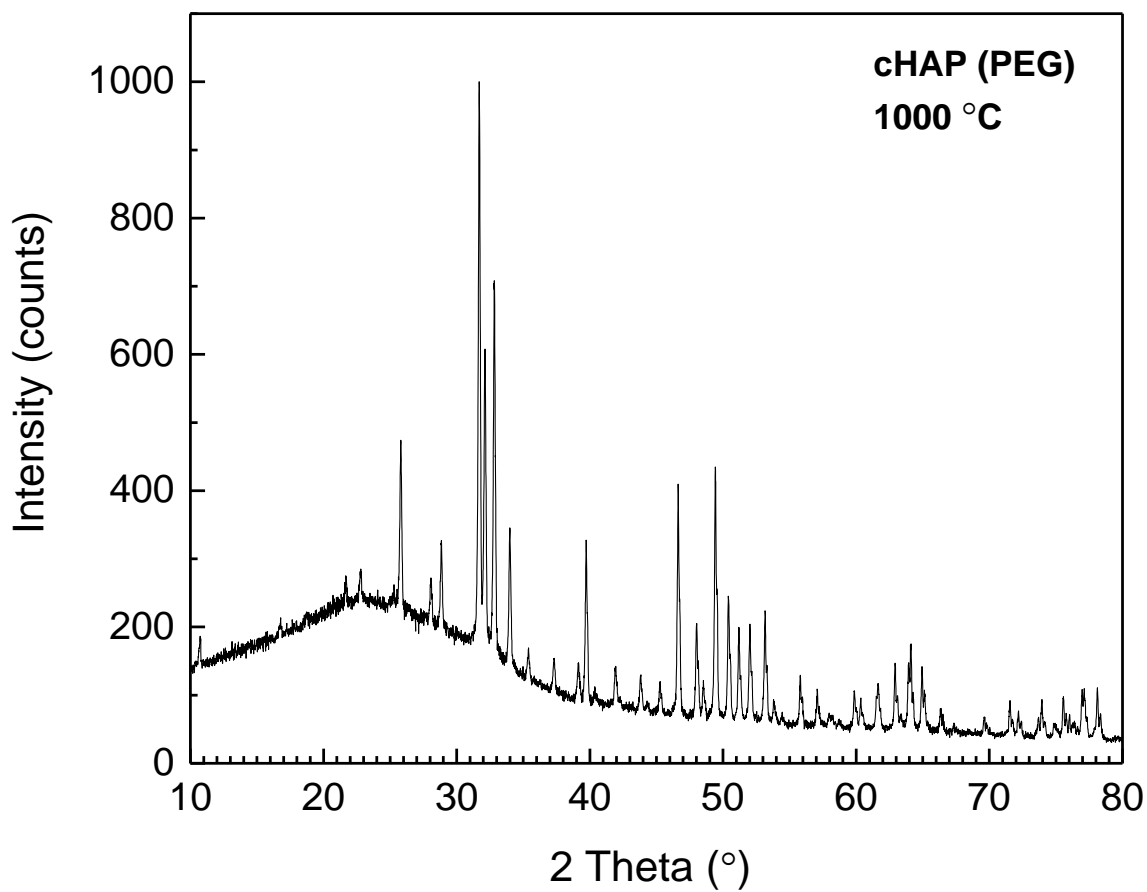

**Figure S3.** XRD diffractogram of cHAP (PEG) annealed at 1000 °C for 5 h (background increment is due to glass sample holder).

**References:**

1. B. Clarke, *Clin. J. Am. Soc. Nephro.*, 2008, **3**, S131-S139.
2. E. Garskaite, K.-A. Gross, S.-W. Yang, T. C.-K. Yang, J.-C. Yang and A. Kareiva, *CrystEngComm*, 2014, **16**, 3950-3959.
3. T. Leventouri, *Biomaterials*, 2006, **27**, 3339-3342.
4. E. Kovaleva, M. Shabanov, V. Putlyaev, Y. Tretyakov, V. Ivanov and N. Silkin, *Cent. Eur. J. Chem.*, 2009, **7**, 168-174.
5. J. E. Barralet, S. M. Best and W. Bonfield, *J Mater Sci.-Mater. M.*, 2000, **11**, 719-724.
6. E. Landi, G. Celotti, G. Logroscino and A. Tampieri, *J. Eur. Ceram. Soc.*, 2003, **23**, 2931-2937.
7. J. P. Lafon, E. Champion and D. Bernache-Assollant, *J. Eur. Ceram. Soc.*, 2008, **28**, 139-147.
8. V. Jokanović, D. Izvonar, M. Dramićanin, B. Jokanović, V. Živojinović, D. Marković and B. Dačić, *J Mater. Sci.-Mater Med.*, 2006, **17**, 539-546.
9. J. M. Antonucci, D. W. Liu and D. Skrtic, *J. Disper. Sci. Technol.*, 2007, **28**, 819-824.

10. T. V. Safronova, V. I. Putlayev, A. V. Belyakov and M. A. Shekhirev, *Mater. Res. Soc. Symp. Proc.*, 2006, **887**, 237-242.
11. S. Morimune-Moriya, S. Kondo, A. Sugawara-Narutaki, T. Nishimura, T. Kato and C. Ohtsuki, *Polym. J.*, 2015, **47**, 158-163.
12. G. S. Han, S. Lee, D. W. Kim, D. H. Kim, J. H. Noh, J. H. Park, S. Roy, T. K. Ahn and H. S. Jung, *Cryst. Growth Des.*, 2013, **13**, 3414-3418.
13. G. H. Nancollas and J. A. Budz, *J. Dent. Res.*, 1990, **69**, 1687-1685.
14. R. Z. LeGeros, S. Lin, R. Rohanizadeh, D. Mijares and J. P. LeGeros, *J. Mater. Sci.-Mater. M.*, 2003, **14**, 201-209.
15. D. Eichert, C. Drouet, H. Sfihi, C. Rey, and C. Combes in *Biomaterials Research Advances*, ed. J. B. Kendall, Nova Science Publishers, Inc., New York, 2007, ch. **5**, pp. 93-143.
16. K. Ishikawa, *Materials*, 2010, **3**, 1138.
17. H. Y. Juang and M. H. Hon, *Biomaterials*, 1996, **17**, 2059-2064.
18. B. Thavornyutikarn, N. Chantarapanich, K. Sitthiseripratip, G. A. Thouas and Q. Chen, *Prog. Biomat.*, 2014, **3**, 61-102.
19. T. Liu, X. Ding, D. Lai, Y. Chen, R. Zhang, J. Chen, X. Feng, X. Chen, X. Yang, R. Zhao, K. Chen and X. Kong, *J. Mater. Chem. B*, 2014, **2**, 6293-6305.
20. R. Govindan, G. S. Kumar and E. K. Girija, *RSC Adv.*, 2015, **5**, 60188-60198.
21. F. S. Senatov, K. V. Niaza, M. Y. Zadorozhnyy, A. V. Maksimkin, S. D. Kaloshkin and Y. Z. Estrin, *J. Mech. Behav. Biomed.*, 2016, **57**, 139-148.
22. B. D. Boyan, T. W. Hummert, D. D. Dean and Z. Schwartz, *Biomaterials*, 1996, **17**, 137-146.
23. K. J. L. Burg, S. Porter and J. F. Kellam, *Biomaterials*, 2000, **21**, 2347-2359.
24. G. Mendonça, D. B. S. Mendonça, F. J. L. Aragão and L. F. Cooper, *Biomaterials*, 2008, **29**, 3822-3835.
25. A. Marino, G. Ciofani, C. Filippeschi, M. Pellegrino, M. Pellegrini, P. Orsini, M. Pasqualetti, V. Mattoli and B. Mazzolai, *ACS Appl. Mater. Interfaces*, 2013, **5**, 13012-13021.
26. R. A. Surmenev, M. A. Surmeneva and A. A. Ivanova, *Acta Biomater.*, 2014, **10**, 557-579.
27. C. Zhao, L. Xia, D. Zhai, N. Zhang, J. Liu, B. Fang, J. Chang and K. Lin, *J. Mater. Chem. B*, 2015, **3**, 968-976.
28. Q. Zhang, H. Dong, Y. Li, Y. Zhu, L. Zeng, H. Gao, B. Yuan, X. Chen and C. Mao, *ACS Appl. Mater. Interfaces*, 2015, **7**, 23336-23345.
29. A. C. De Luca, M. Zink, A. Weidt, S. G. Mayr and A. E. Markaki, *J. Biomed. Mater. Res. A*, 2015, **103**, 2689-2700.
30. K. Kushiro, T. Sakai and M. Takai, *Langmuir*, 2015, **31**, 10215-10222.
31. M. Malinauskas, A. Zukauskas, S. Hasegawa, Y. Hayasaki, V. Mizeikis, R. Buividas and S. Juodkazis, *Light Sci. Appl.*, 2016, **5**, e16133, doi:10.1038/lsa.2016.
32. I. Voynarovych, S. Schroeter, R. Poehlmann and M. Vlcek, *J. Phys. D: Appl. Phys.*, 2015, **48**, 265106.
33. L. Jonušauskas, S. Rekštytė and M. Malinauskas, *Opt. Eng.*, 2014, **53**, 125102-125102.
34. K. A. Gross, V. Gross and C. C. Berndt, *J. Am. Ceram. Soc.*, 1998, **81**, 106-112.

35. K. Tőnsuaadu, M. Peld and V. Bender, *J. Therm. Anal. Calorim.*, 2003, **72**, 363-371.
36. A. Yasukawa, K. Kandori and T. Ishikawa, *Calcified Tissue Int.*, 2003, **72**, 243-250.
37. E. Champion, *Acta Biomater.*, 2013, **9**, 5855-5875.
38. A. L. Boskey and A. S. Posner, *J. Phys. Chem.-US.*, 1973, **77**, 2313-2317.
39. S. Koutsopoulos, *J. Biomed. Mater. Res.*, 2002, **62**, 600-612.
40. I. S. Neira, Y. V. Kolen'ko, O. I. Lebedev, G. Van Tendeloo, H. S. Gupta, F. Guitián and M. Yoshimura, *Cryst. Growth Des.*, 2008, **9**, 466-474.
41. J. C. Merry, I. R. Gibson, S. M. Best and W. Bonfield, *J. Mater. Sci.-Mater. M.*, 1998, **9**, 779-783.
42. M. E. Fleet and X. Liu, *J. Solid State Chem.*, 2004, **177**, 3174-3182.
43. I. Rehman and W. Bonfield, *J. Mater. Sci.-Mater. M.*, 1997, **8**, 1-4.
44. C. Rey, C. Combes, C. Drouet and D. Grossin in *Comprehensive Biomaterials*, ed. P. Ducheyne, K. E. Healy, D. W. Hutmacher, D. W. Grainger and C. J. Kirkpatrick, Elsevier Science, 1st edn., 2011, ch. 1, pp. 187-198.
45. M. Malinauskas, S. Rekštytė, L. Lukoševičius, S. Butkus, E. Balčiūnas, M. Pečiukaitytė, D. Baltriukienė, V. Bukelskienė, A. Butkevičius, P. Kucevičius, V. Rutkūnas and S. Juodkazis, *Micromachines*, 2014, **5**, 839-858.
46. W. Jia, Y. Luo, J. Yu, B. Liu, M. Hu, L. Chai and C. Wang, *Opt. Express*, 2015, **23**, 26932-26939.
47. N. Shi and Q. Dou, *Polym. Composite.*, 2014, **35**, 1570-1582.
48. H. Tsuji and Y. Ikada, *Macromolecules*, 1993, **26**, 6918-6926.
49. A. Shakoor and N. L. Thomas, *Polym. Eng. Sci.*, 2014, **54**, 64-70.